\documentclass{aa}  
\usepackage{graphicx}
\usepackage{txfonts}
\usepackage{hyperref}
\usepackage{natbib}
\bibpunct{(}{)}{;}{a}{}{,} 

\usepackage{subfig}

\begin{document} 
   \title{Radius valley scaling among low mass stars with TESS} 
    \author{
      Harshitha M. Parashivamurthy\inst{1}\thanks{harshitha@das.uchile.cl}
      \and
      Gijs D. Mulders\inst{2}
    }
    
    \institute{
      Departamento de Astronomía, Universidad de Chile, Camino El Observatorio 1515, Las Condes, Santiago, Chile
      \and
      Instituto de Astrofisica, Pontificia Universidad Cat\'olica de Chile, Av. Vicu\~na Mackenna 4860, 7820436 Macul, Santiago, Chile
    }

   \date{Received ; accepted }

  \abstract  
{
The Transiting Exoplanet Survey Satellite (TESS) has been highly successful in detecting planets in close orbits around low-mass stars, particularly M dwarfs. This presents a valuable opportunity to conduct detailed population studies to understand how these planets depend on the properties of their host stars. The previously observed radius valley in Sun-like stars has not been unambiguously detected among M dwarfs, and how its properties varies when compared with more massive stars remains uncertain. We use a volume-limited sample of low mass stars with precise photometric stellar parameters from the bioverse catalog of TESS Objects of Interest (TOIs) confirmed planets and candidates within 120 pc. We detect the radius valley around M dwarfs at a location of 1.64 $\pm$ 0.03 $R_{\oplus}$ and with a depth of approximately 45${\%}$. The radius valley among GKM stars scales with stellar mass as $R_p \propto M_*^{0.15\pm 0.04}$. The slope is consistent, within 0.3$\sigma$, with those around Sun-like stars. For M dwarfs, the discrepancy is 3.6$\sigma$ with the extrapolated slope from the Kepler FGK sample, marking the point where the deviation from previous results begins. Moreover, we do not see a clear shift in the radius valley between early and mid M dwarfs. The flatter scaling of the radius valley for lower-mass stars suggests that mechanisms other than atmospheric mass loss through photoevaporation may shape the radius distribution of planets around M dwarfs. Comparison of the slope with various planet formation and evolution models matches well with pebble accretion models including waterworlds, indicating a potentially different regime of planet formation that can be probed with exoplanets around the lowest mass stars.}


   \maketitle

\section{Introduction}

\hspace{0.3cm}Over the past decade, planet detections through the transit method have significantly increased, facilitated by missions such as the Kepler spacecraft and the Transiting Exoplanet Survey Satellite (TESS) \citep{ricker_2015}. Kepler and its extended mission, K2 \citep[]{Borucki_etal.2010a, batalha2013, Howell_2014, Thompson_2018}, have discovered more than 2,800 confirmed planets, with an additional 3,000 candidates yet to be confirmed. This large dataset enables comprehensive population studies of planetary architectures \citep[e.g.][]{mulders2018,Dattilo_2023}. Demographic studies indicate that more than half of Sun-like star hosts at least one low-mass, short-period planet \citep[]{2010Sci...327..977B,Batalha_2011,he_2019}. The occurrence rates of planets smaller than Neptune are particularly high around low-mass stars \citep[]{Dressing_2013, Dressing_2015,Mulders_2015, hsu2020}.\\

By accurately measuring the stellar parameters, the properties of planets can be measured more precisely, allowing for additional features to be detected. With precise radius measurements from the California-Kepler Survey \citep{Petigura_2017}, a deficit of planets at approximately 1.7 $R_{\oplus}$ around Sun-like stars was observed \citep{Fulton_2017}. This radius valley divides the planets into two groups: super-Earths and mini-Neptunes. This corresponds with an earlier discovery that the densities of small planets are divided into similar categories, with the smaller super-Earths with a rocky core with a thin or no atmospheres and the mini-Neptunes with rocky cores and thick atmospheres \citep[]{Rogers_2015,wolfgang2016}.\\

The radius valley provides insights into planet formation and evolution around various types of stars, with several mechanisms proposed to explain its origin. Photoevaporation, the most accepted process, involves extreme ultraviolet radiation stripping away hydrogen-helium envelopes over time, leading to smaller planetary radii \citep[]{owen&wu2017, owen&clay2018, Wu_2019, rogers2021}. Core-powered mass loss, described by, \cite{ginzburg2018}, \cite{gupta2019}, \cite{gupta2020}, relies on the planet's residual core luminosity and incident bolometric flux to erode its atmosphere over time resulting in bare rocky core planets. However, \cite{tang2024} found that outside the boil-off phase, the core-powered escape is not able to drive significant mass loss.\\

Alternatively, a dichotomy in planet core composition between rocky and water worlds \citep[e.g.][]{Izidoro_2022} could also create a radius valley. Planet formation models integrate pebble accretion with photoevaporation following disk dispersal, accounting for the presence of water worlds to explain the presence of radius valley among low mass stars \citep{venturini2024}. Other planet formation-evolution models have also been put forth to explain the cause of the radius valley, such as, impact erosion \citep{wyatt+2019} and late planet formation in either gas-poor or even gas-empty discs \citep[]{lopez&rice2018, Lee&connors2021}. Among these, photoevaporation and water worlds have emerged as the dominant explanations for planets around solar-mass stars.\\

\cite{Wu_2019} observed that the radius valley among Kepler's FGK stars, shifts to smaller radii as the stellar mass decreases, and found a power law dependence between the two, $R_p$ $\alpha$ $M_{\ast}^{\beta}$, with $\beta \in [0.95, 1.40]$ due to photoevaporation among Kepler planets. This linear dependence for FGKM stars has also been explained by core-powered mass loss as observed by \cite{Berger_2020} through a slope of ${\beta} = 0.26_{-0.16}^{+0.21}$. \cite{bonfanti2023characterisingtoi732bc} observes ${{\partial \log R_{\text{p-valley}}}/{\partial \log M_\star} \approx 0.23}$ and ${0.27}$ for the photoevaporation and core-powered mass-loss models, respectively. The difference between the two inferences from the observations suggests other possible mechanisms that shape the radius valley. \cite{Ho_2023} also notes that thermally driven mass-loss models predict a similar dependence of the valley on stellar mass, based on observations of FGK stars from Kepler. When we extend the sample from FGK to M dwarfs, the planet formation model proposed by \cite{venturini2024} integrating pebble accretion with photoevaporation following disc dispersal, results in a scaling of $R_p \propto M_{\star}^{0.14}$. In this paper, we revisit the radius valley among the TESS planet candidates around GKM stars, to extend the scaling relation of the radius valley to lower mass stars to put tighter constraints on the different proposed hypotheses among the planet size and stellar mass within low mass stars. \\

\begin{figure}[htbp]
        \includegraphics[width=1.1\columnwidth]{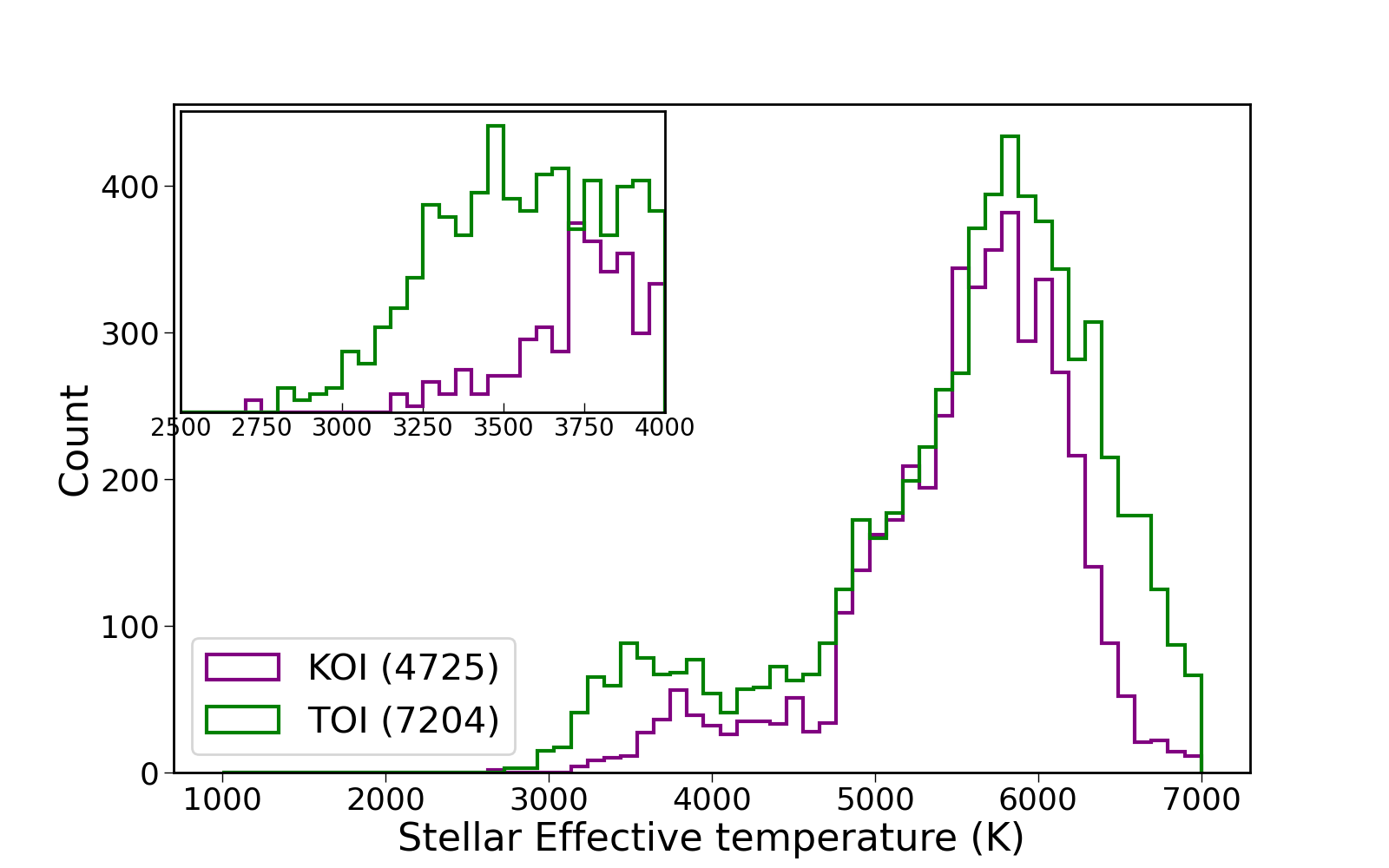}
        \caption{Histogram comparing TESS Objects of Interest (TOIs) and Kepler Objects of Interest (KOIs) from the NASA Exoplanet Archive, with star counts indicated in parentheses. The inset plot (top left) highlights the significant increase in low-mass stars identified by TESS, which enhances our ability to study these stars in greater detail.}
        \label{toivkoi}
    \end{figure}

Despite being optimized for detecting planets around Sun-like stars, the Kepler mission also observed a small number of M dwarfs \citep{Dressing_2013}. This limited sample yielded intriguing results, revealing that M dwarfs host more small transiting planets than Sun-like stars \citep[]{Mulders_2015,Dressing_2015}. However, due to the small sample size (Approximately 85 confirmed Kepler planets around M dwarfs), a larger dataset, such as that provided by TESS, is crucial for better studies. The launch of TESS, with its extensive sky coverage and wide, red optical bandpass filter, is particularly suited to observing M dwarfs \citep[]{Ballard_2019,Barclay_2018}, which are cool and red. Most nearby stars are M dwarfs, presenting a significant opportunity to detect and study planets around the lowest-mass stars (Figure \ref{toivkoi}).\\

With 7341 planet candidates from TESS\footnote{As on November 02, 2024.}, we now have a large dataset for demographic studies. Precise stellar parameters are needed to calculate accurate planet radii and pinpoint the radius valley across different stellar types \citep[e.g.][]{Berger_2020}. M dwarfs, specifically, are known to have a fading radius valley \citep[]{venturini2024, parc2024}, attributing to planet migration. Therefore, we wanted to revisit the radius valley for the M dwarfs, to better constrain planet formation-evolution models.\\ 

\footnotetext{\href{https://exoplanetarchive.ipac.caltech.edu/}{https://exoplanetarchive.ipac.caltech.edu/}}

In this work, we focus on a volume-limited sample of TESS planet candidate hosts with photometric stellar parameters from the bioverse catalogue \citep{bioverse} (Section \ref{lowmasssample}) to calculate their planet radii within 3\% of precision. We employ a Gaussian Mixture Model, GMM (Section \ref{pradii}) to measure the locations of Super-Earth and Mini-Neptune peaks and the radius valley among GKM type stars. Subsequently, we further analyze the shift in the location of the radius valley with stellar mass (Section \ref{scaling}) and discuss the final results and conclusions (Section \ref{conclusions}).
\begin{figure}[htbp]
    \centering
    \includegraphics[width=1\columnwidth]{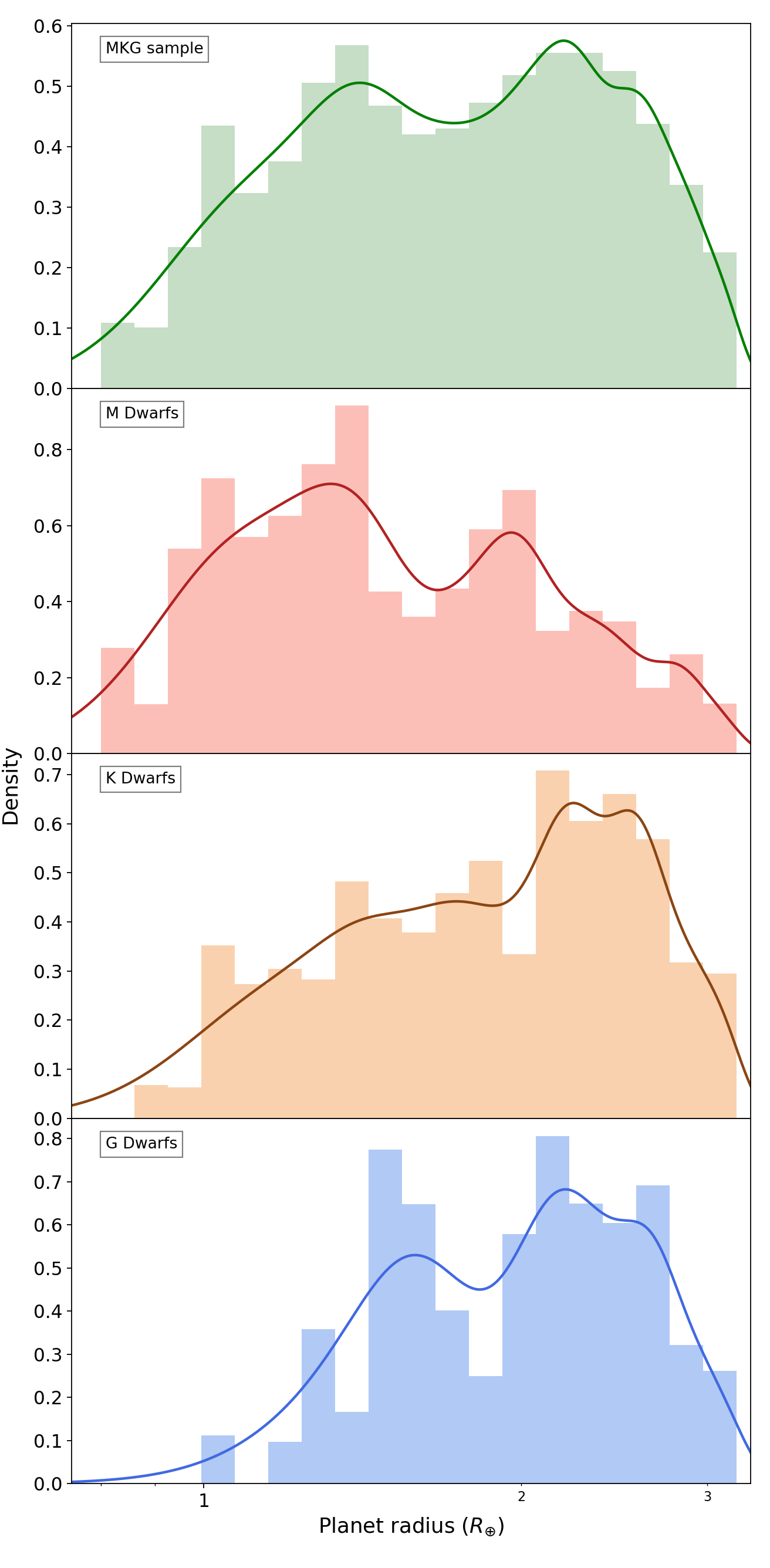}
    \caption{Kernel Density Estimates along with the histograms of Planet radii distribution of the entire sample and M,K,G stellar types.}%
    \label{kdes}
\end{figure}

\begin{figure*}[!h]
    \centering
    \subfloat{\includegraphics[width=0.5\textwidth]{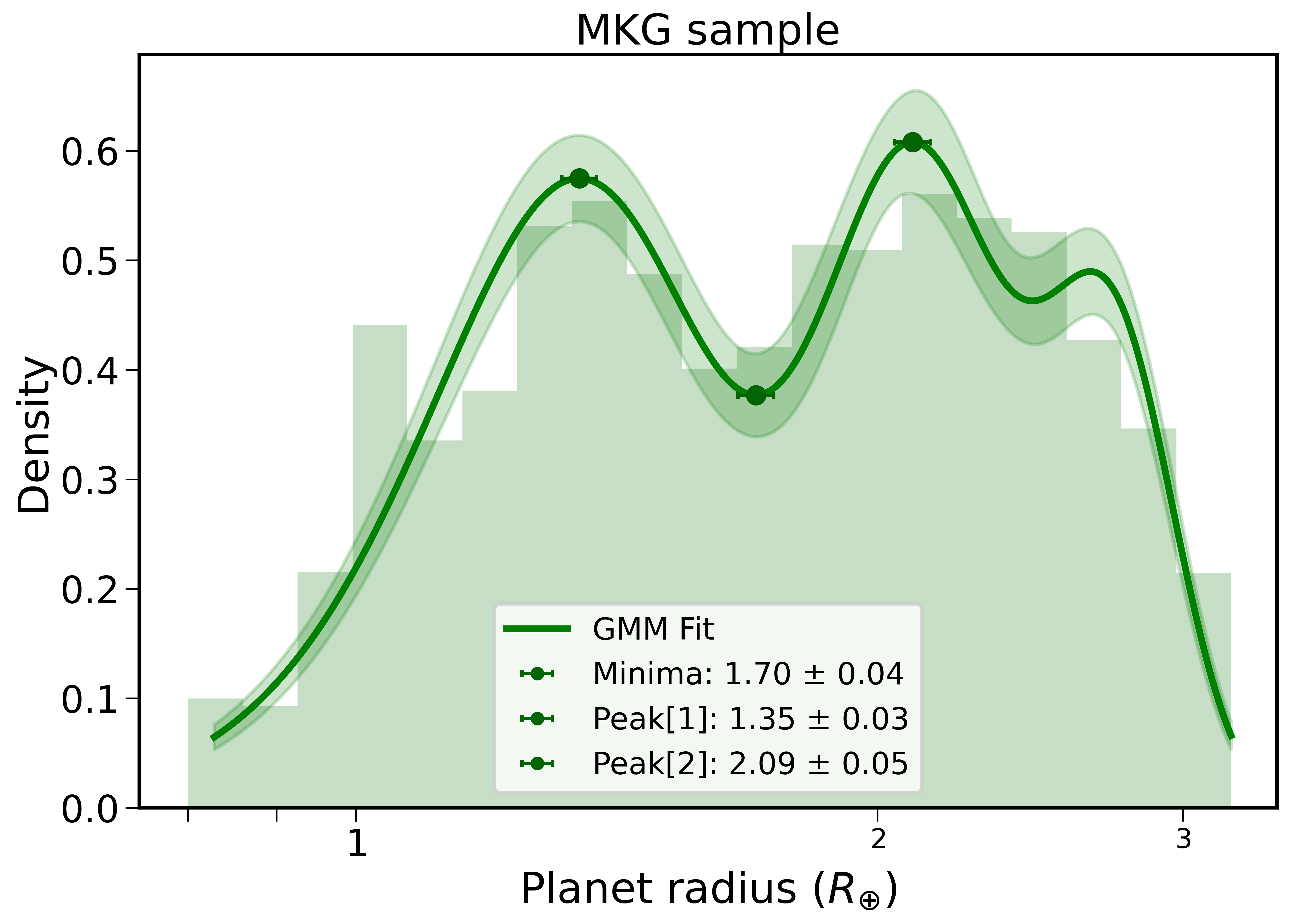}}
    \subfloat{\includegraphics[width=0.5\textwidth]{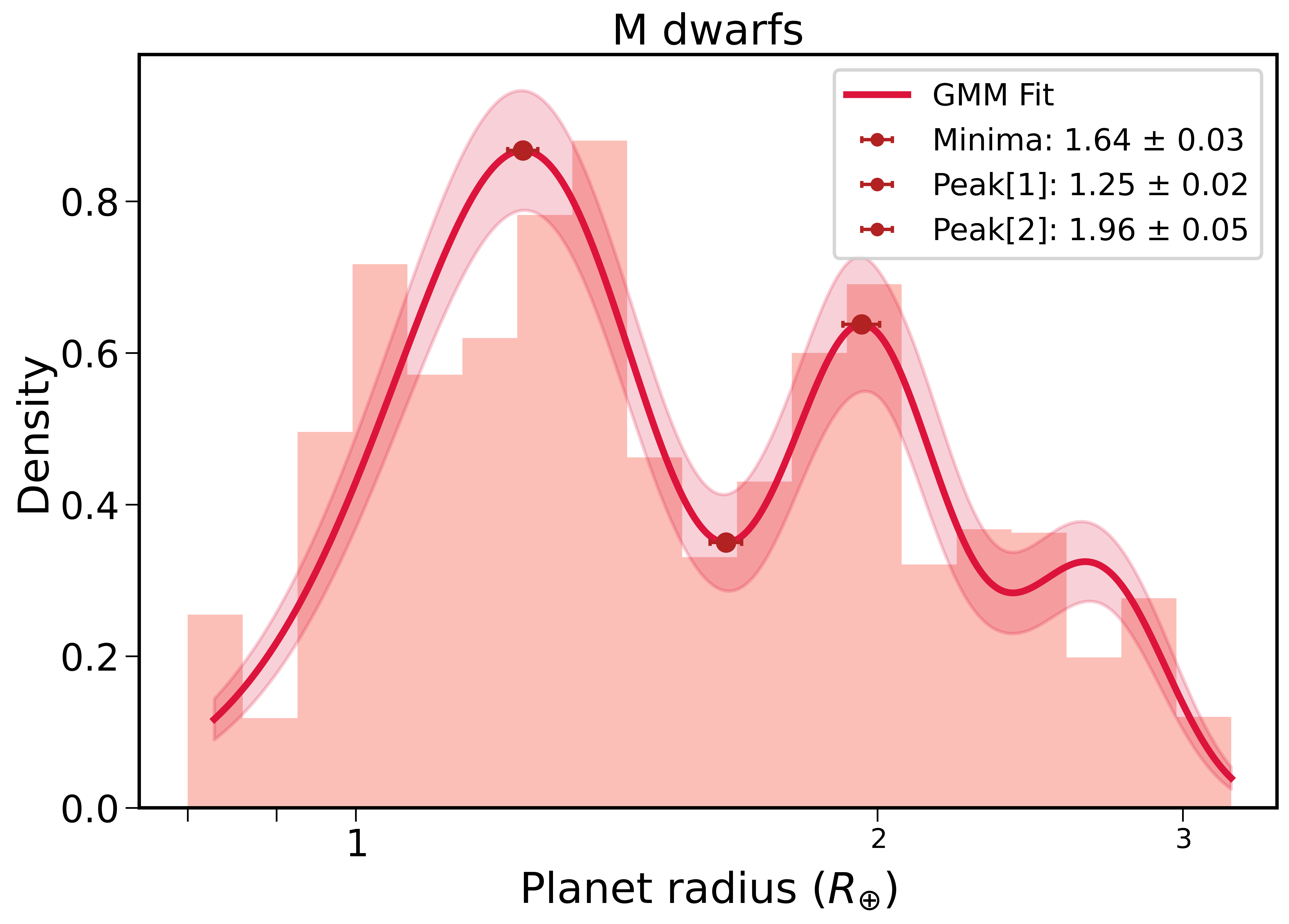}} \\
    
    \subfloat{\includegraphics[width=0.5\textwidth]{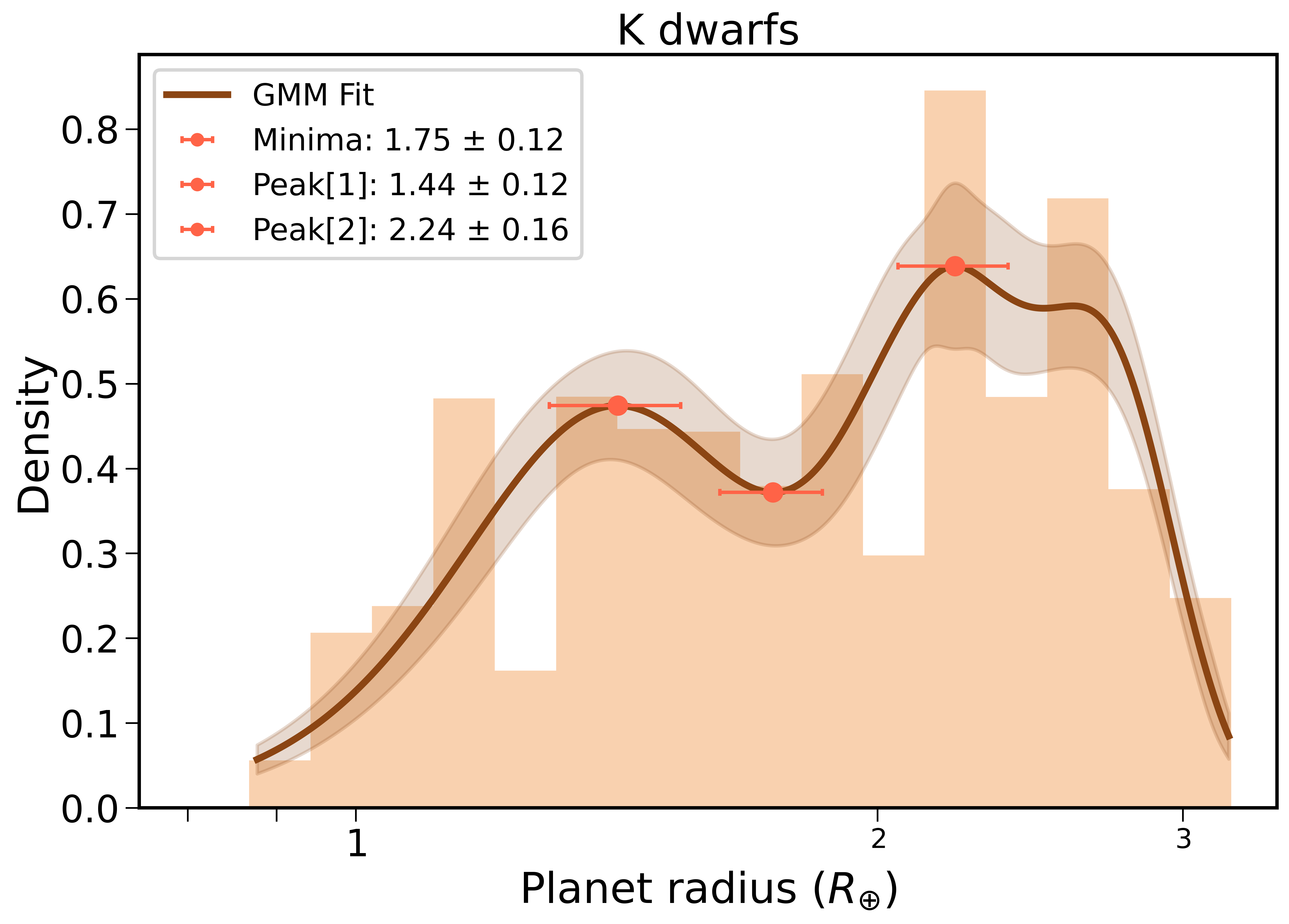}}
    \subfloat{\includegraphics[width=0.5\textwidth]{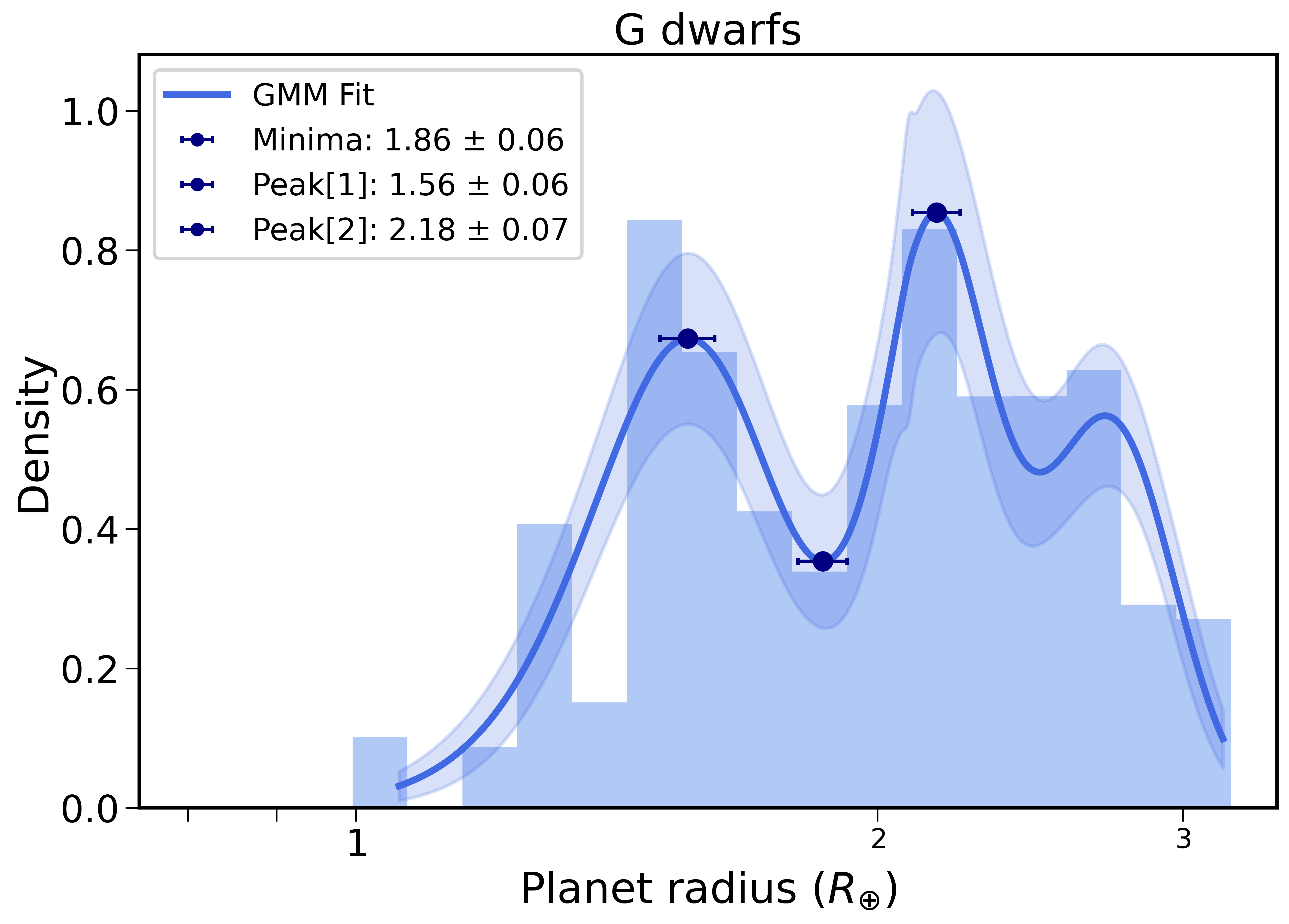}}
    
    \caption{GMM fits to the unbinned radius data, along with corresponding histograms, are presented for the entire low-mass star sample and for M, K, and G stellar types. The Super-Earth and Mini-Neptune peaks, as well as the radius valley, are indicated in each plot.}
    \label{fig:four-subplots}
    \label{gmm_plots}
\end{figure*}

\section{Methods}
\subsection{Low mass star sample}
\label{lowmasssample}

\hspace{0.3cm}For exoplanet demographics, homogeneous stellar samples are needed. The Bioverse, a volume-limited sample (upto 120 pc), based on Gaia DR3 \citep{gaia} parallaxes and photometry, has updated stellar parameters, which we utilized to refine the properties of TESS Input Catalog (TIC) planet hosts \citep[]{tic_2018,Stassun_2019}. The use of the Bioverse catalog, helped bring down the uncertainties to 1\% in effective stellar temperature, 3\% in stellar radius and 5.5\% in stellar mass. \\ 

We cross-matched the coordinates (Right Ascension and Declination) and the GAIA IDs of the stars within 120 pc and with effective temperatures up to 7000 K between the Bioverse and TESS Objects of Interest (TOI) catalogs, resulting in 457 planet candidates and 257 confirmed planets associated with GKM stellar types (Properties of our sample in Table \ref{star_table}). No specific cuts were applied to the sample, other than stellar temperature and distance. We then used these stellar parameters to recalculate the planet radius ($R'_p = R_p * (R'_\star/ R_\star) $). By recalculating the radius of the planet, we reduced the uncertainty of the planet radius from 7.29\% to 3.98\%, which accounts for the transit depth and the stellar radius uncertainty of 2.6\% and 3\% respectively. \\ 

To ensure a sufficiently large number of planets for characterizing the radius valley, we included all the confirmed and candidate planets from the TOI list. We excluded the false positives and false alarms from our sample, using the the Disposition column from the TESS Follow-up Observing Program Working Group (TFOPWG) in the TOI catalog, as most of these were in the 0.5 to 6 $R_\oplus$ range. We also conducted a Triceratops run \citep{triceratops} to assess the likelihood of false positives. The results (Appendix \ref{planets_app}) showed no significant differences between the likely planets and the set of planet candidates identified by Triceratops. Therefore, we chose to proceed with the larger sample that includes planet candidates. While some false positives may still be present in our sample, we prefer the trade-off of having a larger sample size, at the potential costs of a lower reliability, to perform our analysis over a reliable and small sample. This results in a larger sample size, when compared with previous studies of planet radii distributions \citep[]{parc2024,gaidos2024radiusdistributionmdwarfhosted}. In Appendix \ref{planets_app}, we show that when only known and confirmed planets are used for the analysis, the results are consistent within errors; however, by including the candidate planets into our sample, we increase the planet sample for a more robust statistical analysis and reduce errors from the statistics of the small sample.\\


We also ensured that our analysis is not contaminated by binaries. To check for binary stars in our star sample, we looked into the Renormalized Unit Weight Error (RUWE) parameter from the GAIA mission \citep{ruwe-gaia} which is typically used as an indicator of multiplicity. Looking at the distribution of RUWE, stars with values > 1.4 are considered to be possible unresolved binaries. In the bioverse catalog we used, there are a total of 286,391 stars among which only 54,267 stars have a RUWE value greater than 1.4. So 19\% of the stars in the bioverse catalog could be unresolved binaries. In the Bioverse-TOI sample we identify 64 possible binaries, with a 9\% probability. This lower fraction likely means binaries have partly been eliminated during the vetting process. We keep possible binaries in our sample but verified throughout the paper that omitting them did not significantly impact the analysis of the planet radius distribution. \\

\begin{table}[h!]
    \renewcommand{\arraystretch}{1.3} 
    \centering
    \begin{tabular}{lccc}
    \hline
    \hline
    Stellar type          & M dwarfs & K dwarfs & G dwarfs \\
    \hline
    Effective temperature (K) & 3441 & 4605 & 5668 \\
    Stellar mass ($M_\odot$)   & 0.4  & 0.68 & 0.96 \\
    Stellar radii ($R_\odot$)  & 0.4  & 0.71 & 0.98 \\
    TESS magnitude         & 11.76 & 9.98 & 8.86 \\
    \hline
    \end{tabular}
    \caption{Summary of the stellar properties (median values) included in this study.}
    \label{star_table}
\end{table}

Table \ref{mrt} shows our sample, from this dataset, we divided the 843 planet candidates into three temperature bins corresponding to stellar types: M dwarfs (Teff < 3880 K), K dwarfs (Teff < 5340 K), and G dwarfs (Teff< 6040 K). This categorization resulted in 327 M dwarfs, 304 K dwarfs, and 165 G dwarfs. The relatively low number of G dwarfs is a result of the volume limit of Bioverse catalog.

\subsection{Planet radii distribution}
\label{pradii}

\hspace{0.3cm}Figure \ref{kdes} shows the planet radii distribution of the 667 GKM dwarfs hosting planets between 1 to 3 $R_{\oplus}$ (After excluding the False Alarms and False Positives from the TESS Follow-up Observing Program Working Group (TFOPWG) Disposition column). The radius valley around M dwarfs and G dwarfs are clearly observed at 1.63 $R_{\oplus}$ and 1.86 $R_{\oplus}$. Around K dwarfs, the radius valley at 1.76 $R_{\oplus}$ is less prominent, possibly due to the relatively low number of super-Earths compared to the number of mini-Neptunes. Because the prominence of the radius valley in the histograms can depend on the choice of bins, we use a KDE and GMM approach and recover the radius valley.\\

We note that our sample has not been corrected for completeness. Completeness correction can change the relative height of the Super-Earth and Mini-Neptune peaks, thereby shifting the inferred value of the radius gap to lower or upper values. So, we used the \cite{Fulton_2017} dataset to assess the impact of completeness correction on the radius gap and found that the effect was within uncertainties. Therefore, we proceeded with the raw sample, which includes both confirmed and candidate planets.\\

\subsubsection{Kernel Density Estimates (KDEs)} 
Figure \ref{kdes} shows the KDEs of the planet radii distribution along with their histograms revealing the radius valley. Due to the 3.98{\%} uncertainty in the radius measurements, the histogram binning had to be carefully adjusted to accurately capture the radius gap. By adjusting the bandwidth from 0.2 to 0.3, we selected the KDEs that best represent the data. The KDEs helped in representing the bimodality in the planet radii distribution of the small sample space of M,K,G stellar types.\\
 
We observe a subtle valley in both the histogram and KDE around 1.7 $R_{\oplus}$ for the entire sample, consistent with the known bimodal planet radius distribution \citep{Fulton_2017}. The radius valley is not deep, with a minimum of 45$\%$, consistent with shape of the valley in a more heterogeneous sample of host stars. The absence of a well-defined deep radius gap among the planets can be attributed to low number statistics and possible contamination of the sample with false positives. Similarly, radius valleys are clearly observed for the individual G, K, and M stellar spectral types. M dwarfs exhibit a distinct valley, with a small sub-Neptune peak, consistent with previous studies. For K dwarfs, despite the sample containing around 300 planets, the radius valley is not as clearly seen in the histogram compared to M dwarfs, mostly due to a relatively low number of detected super-Earths compared to mini-Neptunes. G dwarfs, however, present a clearer distinction between mini-Neptunes and super-Earths, with evident support for the radius valley.\\

\subsubsection{Gaussian Mixture Models (GMMs)} 
To accurately estimate the location of the radius valley, we used the Gaussian Mixture Model (GMM) from the \texttt{sklearn.mixture} package. GMM is a probabilistic model that assumes data points are generated from a mixture of a finite number of Gaussian distributions with unknown parameters. We implement the GMM fit to the planet radii values in the range, 0.8 - 3.2 $R_\odot$ and evaluated the fit of the GMM, using the Bayesian Information Criterion (BIC), to help determine the goodness of fit. The bootstrapping resampling method was used to account for uncertainties in the data by generating multiple versions of the radius distribution and seeing how sensitive the GMM's features are to the omission of specific data points.\\

The radius valleys  for the stellar types are presented in figure \ref{gmm_plots}. We explored a different number of components (2,3, and more), and found that a three component model provided a better fit to the region around the radius valley (refer Appendix B \ref{gmm_app}). In all cases, the location of the first minimum aligns well with the radius valley seen in the histograms and KDEs. The derived locations and their confidence intervals are listed in Table \ref{gmm} . The valley shifts to smaller radii for later spectral types, a trend we will explore further in the next section.\\

\begin{table*}[!ht]
\small
    \renewcommand{\arraystretch}{2} 
    \centering
    \begin{tabular}{llccccccccc}   
    \hline
    \hline
    GAIA ID & TIC ID 
    &\shortstack{Planet \\ \#} 
    &\shortstack{ $T_{\mathrm{eff}}$\\ (K)} 
    &\shortstack{ $R_\star$ \\ ($R_\odot$)} 
    & \shortstack{$M_\star$\\  ($M_\odot$)} 
    & \shortstack{Distance \\(pc)} 
    & \shortstack{Period \\ (days)} 
    &\shortstack{ $R_p$\\  ($R_\oplus$)} 
    & \shortstack{Planet \\ Disposition }
    & RUWE\\
        \hline
 2.738034573094368e+18 & 331691841 & 1 & 5465 & 0.851 & 0.859 & 57.25 & 6.9341584 & 2.111 & PC & 0.88\\
3.472285521841332e+18 & 107018378 & 1 & 3308 & 0.219 & 0.186 & 36.7 & 3.7924221 & 1.563 & PC & 1.27\\
4.866555051425383e+18 & 77253676  & 1 & 5533 & 0.926 & 0.915 & 92.92 & 8.6078603 & 2.506 & APC & 0.99\\
2.3182959791265e+18  & 251848941 & 1 & 4332 & 0.675 & 0.65  & 62.89 & 6.5578513 & 2.603 & CP & 1.55\\
2.3182959791264998e+18  &251848941 &3 &4332 &0.675 &0.650  &62.890    &9.9615672  &2.198 &CP & 1.55\\
2.3182959791264998e+18  &251848941 &4 &4332 &0.675 &0.650  &62.890    &3.2384253  &1.547 &KP & 1.55\\
5.6119123689978737e+18  &108645766 &1 &3293 &0.307 &0.285  &55.800    &2.9741750  &3.818 &FP & 1.14\\
1.6288406502203866e+18  &198211976 &1 &3584 &0.519 &0.521  &49.590    &0.4018586  &0.551 &FA & 1.30\\

        \hline
    \end{tabular}
    \textbf{\caption{Stellar and planet properties of all the targets in our sample. The table consists stellar parameters from \cite{bioverse}, including the Gaia Renormalized Unit Weight Error (RUWE) values, recalculated planet radius and the planet orbital period and planet disposition from NASA exoplanet archive. The planet disposition is the TESS Follow-up Observing Program Working Group (TFOPWG) Dispostion, where; APC=ambiguous planetary candidate, CP=confirmed planet, FA=false alarm, FP=false positive, KP=known planet, PC=planetary candidate.}
    \label{mrt}
    (The entire table is available in a \href{mrt_table.mrt}{machine readable form.)}}
\end{table*}

\begin{table*}[!ht]
   \renewcommand{\arraystretch}{1.5}
    \centering
    \begin{tabular}{lccc}
        \hline
        \hline
         & \shortstack{Super-Earth Peak \\ ($R_\oplus$)}
         & \shortstack{Radius Valley Minima \\ ($R_\oplus$)}
         & \shortstack{Mini-Neptune Peak \\ ($R_\oplus$)} \\
        \hline
        \multicolumn{4}{@{}l}{\textbf{2 Components}} \\
        GKM Sample    & 1.49 $\pm$ 0.04  & 1.93 $\pm$ 0.04  & 2.42 $\pm$ 0.03 \\
        M Dwarfs      & 1.29 $\pm$ 0.07  & \textbf{1.7 $\pm$ 0.14}  & 2.17 $\pm$ 0.15 \\
        K Dwarfs      & 1.58 $\pm$ 0.18  & \textbf{1.90 $\pm$ 0.06} & 2.46 $\pm$ 0.04 \\
        G Dwarfs      & 1.76 $\pm$ 0.22  & \textbf{2.05 $\pm$ 0.16} & 2.44 $\pm$ 0.16 \\
        \hline
        \multicolumn{4}{@{}l}{\textbf{3 Components}} \\
        GKM Sample    & 1.34 $\pm$ 0.03  & 1.70 $\pm$ 0.04  & 2.09 $\pm$ 0.05 \\
        M Dwarfs      & 1.25 $\pm$ 0.02  & \textbf{1.64 $\pm$ 0.03} & 1.96 $\pm$ 0.04 \\
        K Dwarfs      & 1.43 $\pm$ 0.12  & \textbf{1.75 $\pm$ 0.11} & 2.23 $\pm$ 0.16 \\
        G Dwarfs      & 1.56 $\pm$ 0.05  & \textbf{1.86 $\pm$ 0.06} & 2.17 $\pm$ 0.06 \\
        \hline
    \end{tabular}
    \caption{Super-Earth peaks, Mini-Neptune peaks and radius valley values from 2- and 3-component GMM fits for the entire sample and M, K, and G dwarfs.}
    \label{gmm}
\end{table*}

\section{Planet size scaling with stellar mass}
\label{scaling}
\begin{figure}%
    \subfloat{{\includegraphics[width=9cm]{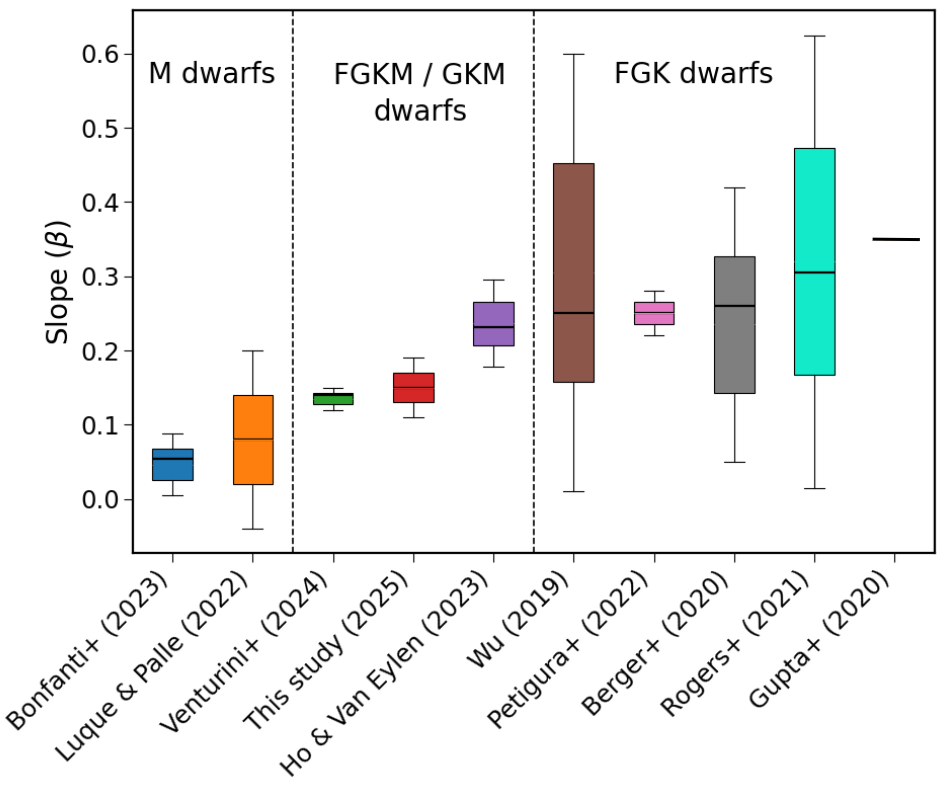} }}%

    \caption{Slope values, $\beta = \left( \frac{\partial \log R_{\text{gap}}}{\partial \log M_*} \right)$ from various studies, grouped by stellar mass ranges (M dwarfs, FGKM dwarfs, FGK dwarfs) in the radius- stellar mass plane. The colored boxes represent the uncertainty ranges, while the horizontal dashes indicate the slope values for each study. A rise in the slope values can be observed as we go from M dwarfs to Sun-like stars.}
    
    \label{obs&the}
\end{figure}

We performed a non-linear least squares fit to quantify the relationship between planet size and stellar mass, observed through the Gaussian Mixture Model (GMM) analysis of the G,K,and M stars. The shift of the radius valley to smaller radius values along decreasing stellar masses, gave us a power-law scaling of $R_{\mathrm{valley}} = 1.87 * M_{\star}^{\beta}$, ${\beta}={\mathrm{0.15\pm0.04}}$. \\

The observed shift of the radius valley with stellar mass aligns with prior observational studies of Kepler and TESS planets (Figure \ref{obs&the}). Studies, such as \cite{Wu_2019}, who found $R_p \propto M_{\star}^{0.25}$ and \cite{Ho_2023}, who reported $\beta = 0.23_{-0.064}^{+0.053}$ for FGK stars. Similarly, \cite{Berger_2020} obtained $\beta = 0.26_{-0.16}^{+0.21}$ and \cite{Petigura_2022} identified a  slope of $\beta = 0.18_{-0.07}^{+0.08}$ for FGKM stars. While a more recent study by \cite{berger2023} determined a slope of $\left( \frac{\partial \log R_{\text{gap}}}{\partial \log M_*} \right)_{S} = -0.046_{-0.117}^{+0.125}$ under constant incident flux (S) for Kepler planets. However, for low-mass stars, \cite{Luque_2022} observed a slope of ${\beta} = {0.08 \pm 0.12}$ and \cite{bonfanti2023characterisingtoi732bc}, ${\beta} = 0.054 _{-0.034}^{+0.049}$  for M dwarfs, within TESS planets. \\

While the inferred slope among GKM stars agrees well with that initially inferred from Kepler by  \citet{Wu_2019}, the results for M dwarfs are different, a discrepancy of approximately 3.6$\sigma$ is observed for the M dwarf data point, pointing to a distinct planet formation and evolution mechanism for these stars (Figure \ref{5points}). We will characterize the radius valley scaling among M dwarfs in more detail in the next section.

\subsection{Radius valley among the M dwarfs}
\begin{figure}%
    \subfloat{{\includegraphics[width=8cm]{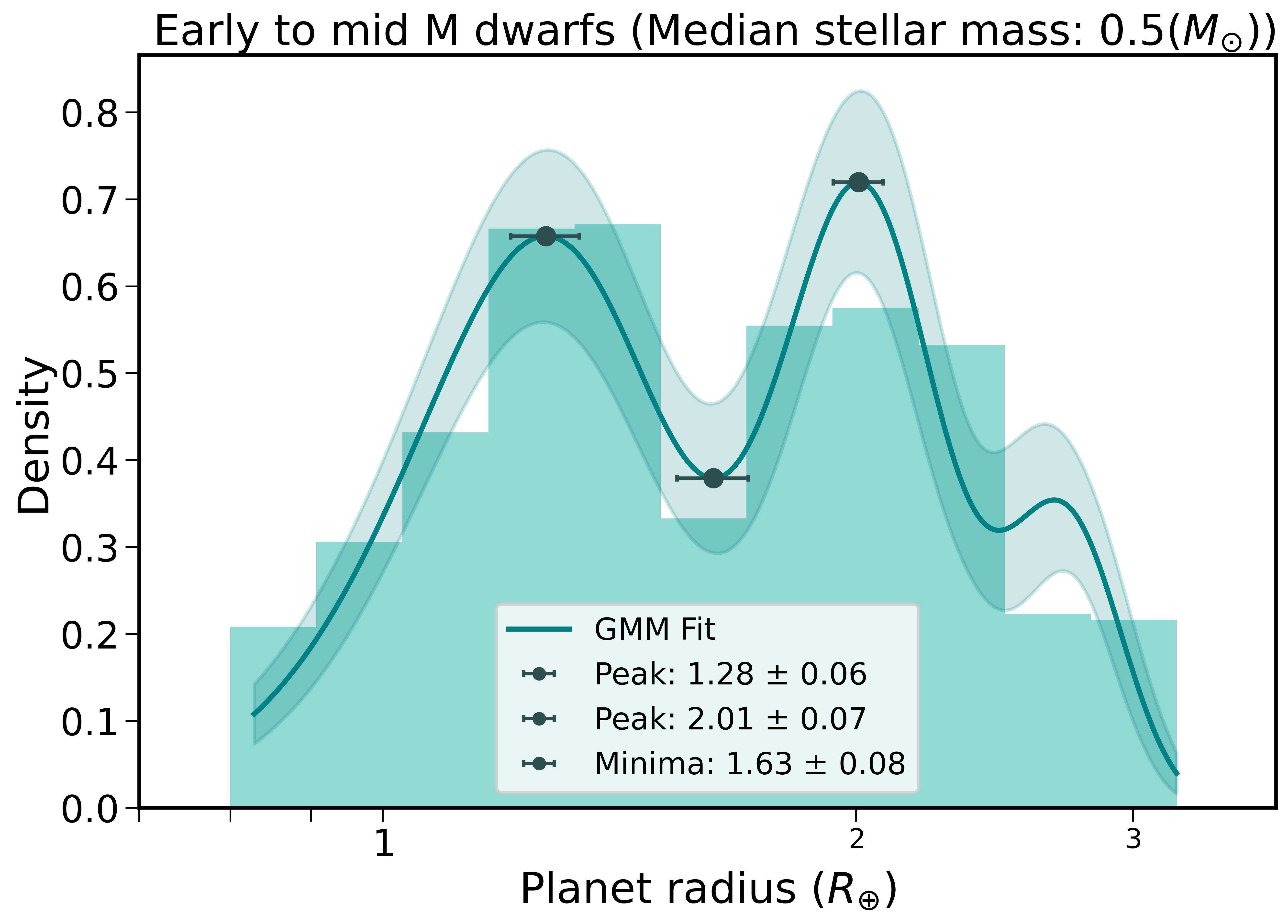} }}%
    
    \subfloat{{\includegraphics[width=8cm]{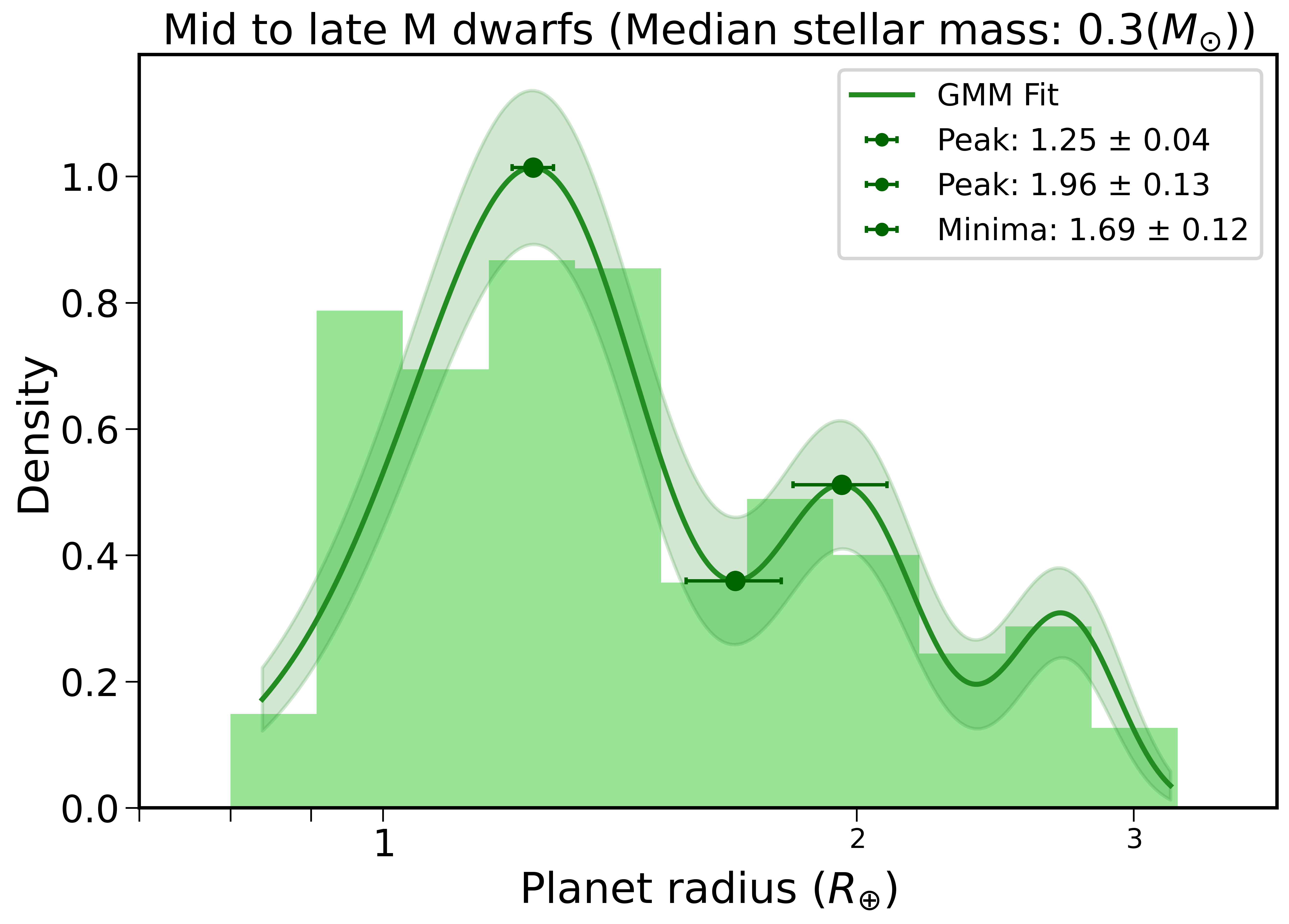} }}%
    
    \caption{GMM fits along with their histograms for early and late M dwarfs.}%
    \label{earlyandlate}
\end{figure}

In our dataset, we identified approximately 320 M dwarfs with effective temperatures ranging from 2637 to 3870 K, which we here divide into two groups to create early and late M dwarf samples. The early M dwarf group (M0-M3, 3440-3880 K) contains 163 stars, with a median 0.502$M_\star$. Similarly, the late M dwarf group (M4-M8, 2630-3440 K) also includes 164 stars, with a median mass of 0.299$M_\star$. Following the same procedure outlined in Section \ref{lowmasssample}, we plotted a histogram along with the GMM fit. We observe a clear radius valley at 1.63 $\pm$ 0.08 $R_\oplus$ and 1.69 $\pm$ 0.12 $R_\oplus$ for the early and late M dwarf samples, respectively (Figure \ref{earlyandlate}). The scaling is, within the uncertainties, consistent with $\beta$ = 0, indicating no evidence for a scaling relationship. However, the uncertainties are sufficiently large that the existence of a scaling cannot be ruled out. Additionally, no evidence for a slope is observed among the M dwarfs.\\

\begin{figure}[htbp]
        \includegraphics[width=1.1\columnwidth]{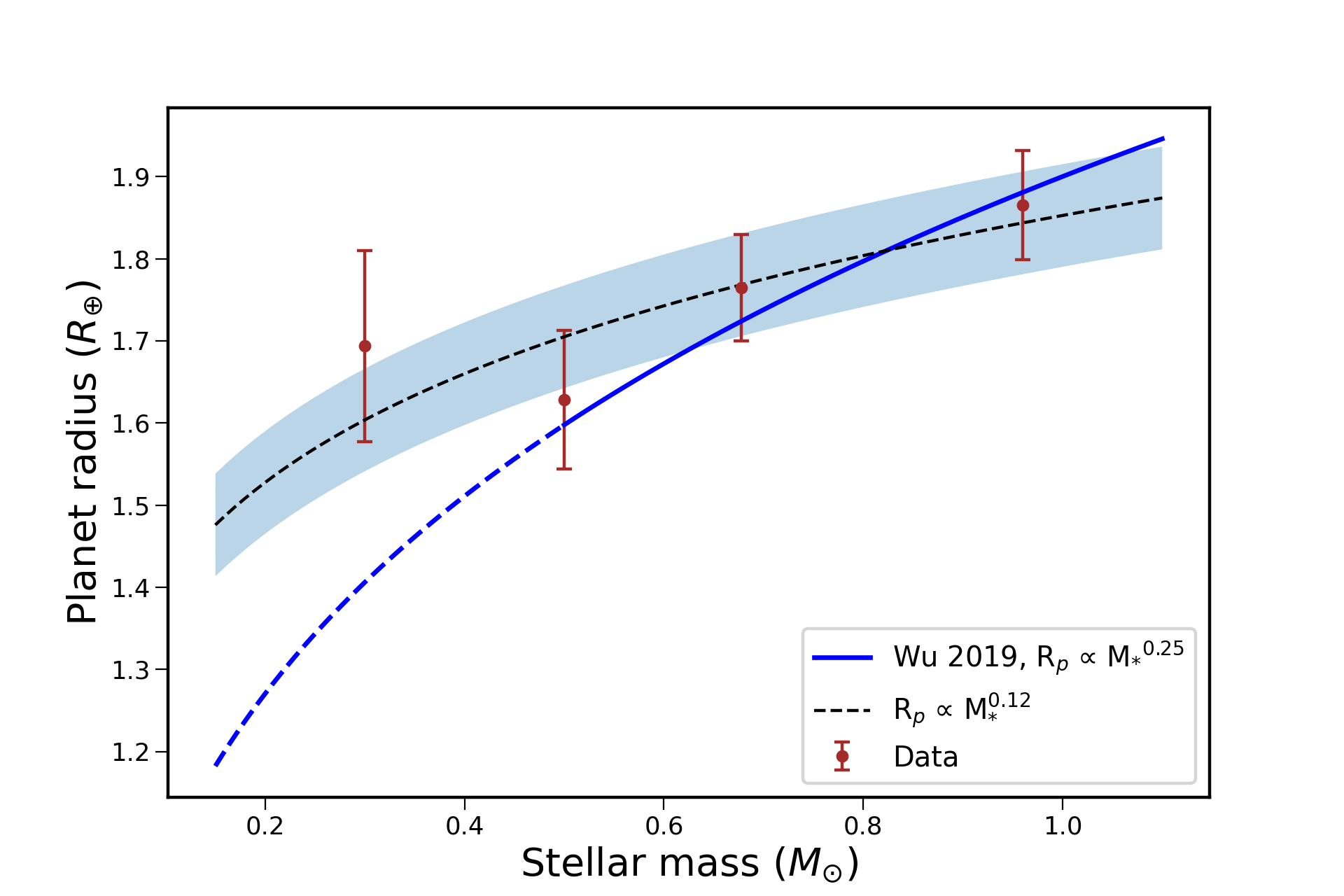}
        \caption{A comparison of our data with that of \cite{Wu_2019}. The scaling from Wu is plotted with an offset for better alignment and easier comparison.}
        \label{5points}
    \end{figure}

With the new radius valley values for the early and late M dwarfs, we updated our power-law scaling with these additional points, and found a shallower slope of, $\beta$ = 0.12 $\pm$ 0.06. Here, we can observe that the new data points are consistent with our previous fit, where the slope gets less steep when compared with FGK stars (Figure \ref{5points}). The comparison with \cite{Wu_2019} shows that the slope matches for the Sun-like stars but when M dwarfs are included, a diversion is observed. This deviation in the slope of the radius valley may suggest a break in the trend predicted by photoevaporation models for FGK stars. For M dwarfs, this deviation may point to the need for additional mechanisms, such as the inclusion of water worlds or pebble accretion models. The implications of this deviation will be discussed further in Section \ref{conclusions}.\\

\section{Conclusions and Discussions}
\label{conclusions}
We created a volume-limited sample by crossmatching 7341 TESS project candidates from the TESS exoplanet archive\footnote{\href{https://exoplanetarchive.ipac.caltech.edu/}{https://exoplanetarchive.ipac.caltech.edu/}} with the Bioverse catalog for better precision of stellar effective temperatures and a 3\% accurate planet radius measurements of all the TESS Objects of Interests(TOIs). With this refined sample of 843 planets, we examined the radius valley among the GKM stellar types with histograms, Kernel Distribution Estimates (KDE) and Gaussian Mixture Models (GMM). We use GMM to measure the location of the radius valley and its associated uncertainty for spectral types M, K, and G. Our main conclusions are: \\
\begin{enumerate}
  \item A clear radius valley is observed among the 327 M dwarfs at a radius of 1.64 ${\pm}$ 0.03 $R_{\oplus}$ and a depth of 45$\%$. Splitting the M dwarf sample into mid and late M dwarfs, with median masses, 0.30 and 0.502 $M_{\odot}$ respectively, we find the radius valleys at 1.63 ${\pm}$ 0.08 $R_{\oplus}$ and 1.69 ${\pm}$ 0.12 $R_{\oplus}$ respectively. 
  
  \item We find that the location of the radius valley increases with spectral type, ranging from 1.63 $\pm$ 0.03 $R_{\oplus}$ for M dwarfs to 1.86 $\pm$ 0.06  $R_{\oplus}$ for G dwarfs. By fitting a scaling relation with stellar mass for GKM stars, we derive $R_p = M_{\star}^{0.15 \pm 0.04}$. This scaling is shallower compared to that observed for Kepler FGK stars.

  \item We detect no radius valley scaling with stellar mass among M dwarfs, possibly due to a small sample size (Figure \ref{5points}). In addition, the location of the valley in M dwarfs is significantly higher than expected based on extrapolating the slope around Kepler FGK stars at 3.6$\sigma$.

  \item The steep slope around FGK stars matches well with atmospheric mass loss through photoevaporation. However, the fltattening of the slope around M dwarfs may indicate a different mechanism may play a role for lower mass stars. Our observations match particularly well with the pebble accretion models including water worlds from \cite{venturini2024}.
    
\end{enumerate}

The relatively larger homogeneous dataset with updated stellar parameters allowed us to clearly identify the radius valley among M dwarfs—a feature that many prior studies missed. Previous studies with datasets less than 200 planets with mass measurements did not observe this feature clearly or rather observed a fading radius valley among the low mass stars \citep[e.g.][]{Luque_2022, parc2024}. \\

The observed power-law scaling from this study, $R_{p}$ ${\alpha}$ $M_{*}^{0.15}$, aligns well with previous models and observation studies conducted for Kepler and TESS planets. The slopes obtained for FGK stars, by \cite{Wu_2019} seems consistent with photoevaporation. \cite{berger2023} finds that core-powered mass-loss dominates over photoevaporation in shaping the radius valley among Kepler planets. \cite{Petigura_2022} also found a power-law scaling for FGKM stellar types in support of mass-loss mechanisms. Interestingly, \cite{rogers2021} found either photoevaporation and core-powered mass-loss models consistent with their two data sets of California-Kepler Survey and Gaia-Kepler Survey. However, when M dwarfs are included, the slope seems to get shallower as seen in this study and others (Figure \ref{obs&the}) \citep[]{Luque_2022, bonfanti2023characterisingtoi732bc}.\\

The flatter slopes around M dwarfs are predicted by certain planet formation and evolution models. For example, \cite{venturini2024}, which uses pebble accretion models to create a population of rocky planets and water worlds, predict a slope of $R_{p}$ ${\alpha}$ $M_{*}^{0.14}$. Therefore, it is evident that as we shift towards the lowest-mass stars, additional mechanisms such as pebble accretion or the inclusion of water worlds may be necessary to account for the relatively flat slopes, while photoevaporation alone maybe insufficient.\\ 

A limitation of this analysis is that it is based on a sample of planet candidates that may include an unknown number of false positives. While a large sample size is essential to robustly detect the radius valley across different spectral types, ongoing efforts to validate candidates and remove false positives may influence the results. However, a comparison between confirmed planets and candidates showed no significant differences, so we proceeded with the full sample including candidates.\\

To evaluate the impact of completeness corrections, we performed a comparison using the \citet{Fulton_2017} dataset. We examined the radius valley before and after applying a completeness correction; although the valley remains visually consistent in the histogram, the GMM analysis reveals a slight shift across all stellar types. This shift likely arises from differing completeness corrections between super-Earths and mini-Neptunes, but the planet radius–stellar mass slope remains unaffected. While our volume-limited sample improves homogeneity, we did not apply a completeness correction near the detection threshold. This omission may particularly affect the radius valley detection for K dwarfs, where super-Earths appear underrepresented relative to mini-Neptunes. A more rigorous completeness treatment is needed to refine these results and will be pursued in future work.\\

\begin{acknowledgements}
The authors thank the referee for providing insightful comments, that improved this work and made the results clearer. H.M.P acknowledges support from ANID (Beca de doctorado nacional) folio de postulacion 21241689 and Millennium Institute of Astrophysics (MAS), and would like to thank the thesis committee members for their valuable advice over time. G.D.M. acknowledges support from FONDECYT project 11221206, from ANID---Millennium Science Initiative---ICN12\_009. This research has made use of the NASA Exoplanet Archive, which is operated by the California Institute of Technology, under contract with the National Aeronautics and Space Administration under the Exoplanet Exploration Program. This paper includes data collected by the TESS mission. Funding for the TESS mission is from the NASA Science Mission directorate. The results reported herein benefited from collaborations and/or information exchange within NASA’s Nexus for Exoplanet System Science (NExSS) research coordination network sponsored by NASA’s Science Mission Directorate and project “Alien Earths” funded under Agreement No. 80NSSC21K0593. We also want to thank the \textit{TESS} team at MIT for all the work done in making the mission happen and the CALTECH team for maintaining the NASA Exoplanet Archive. 
\end{acknowledgements}

\bibliographystyle{aa}
\bibliography{sources}

\begin{appendix}

\section{Gaussian Mixture Model (GMMs)}
\label{gmm_app}
\begin{figure}[htbp]
        \includegraphics[width=1\columnwidth]{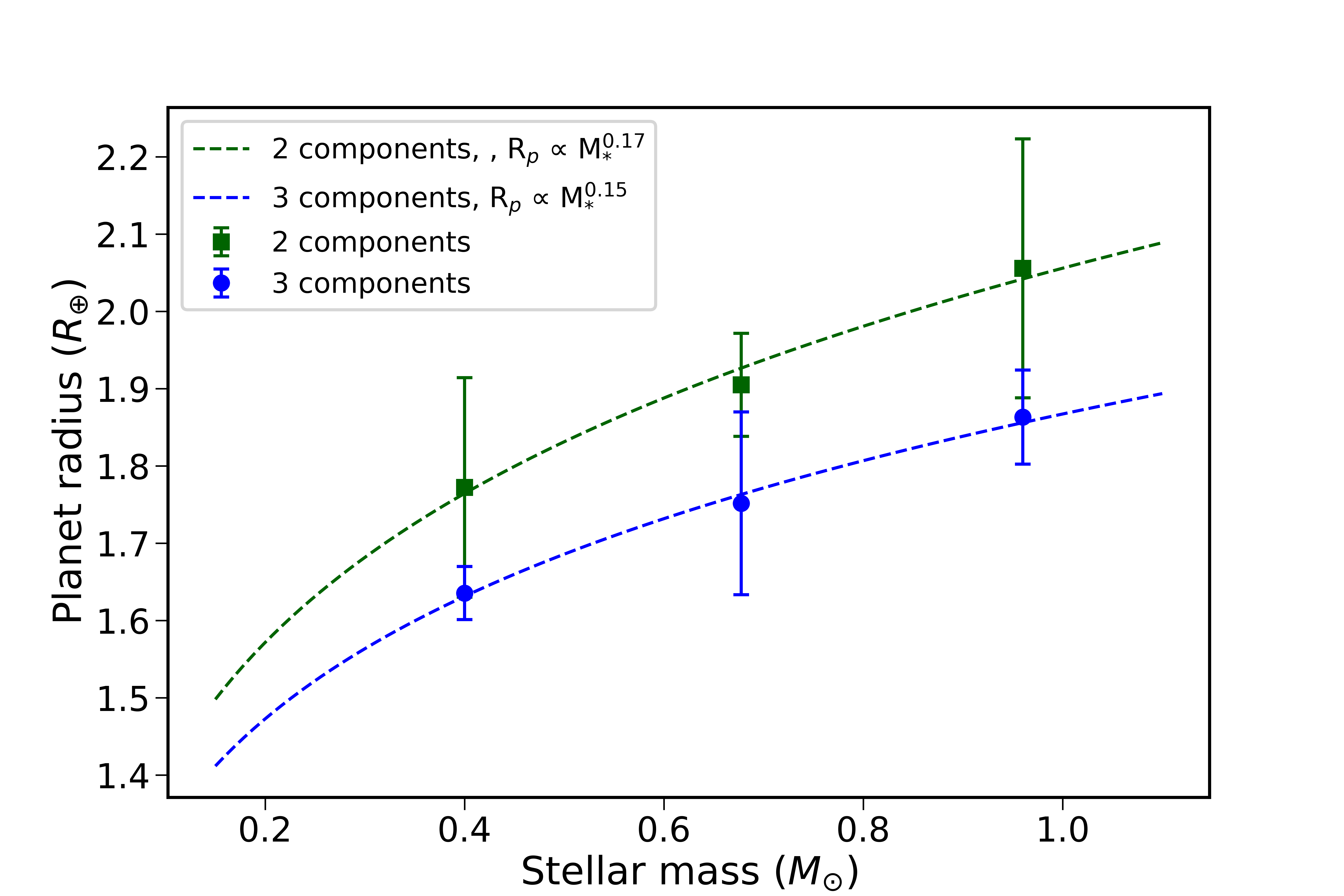}
        \caption{Comparison of the 2- and 3-component GMM fits for the radius valley values reveals that the inclusion of a third component improves the fit.}
        \label{2and3}
    \end{figure}


To obtain the accurate values of the radius valley that we observed in the histograms and KDE, we utilized Gaussian Mixture Modeling, a package from \texttt{sklearn.mixture}. GMM is a probabilistic model that assumes that all the data points are generated from a mixture of a finite number of Gaussian distributions with unknown parameters. Therefore, we obtain the precise values of the bimodality, such as the Peaks and minima of the fit. The bootstrapping resampling method was used to account for uncertainties and to estimate the confidence intervals for the peaks and minima locations, providing more robust results than fitting a single GMM on the original data. \\

\begin{figure}[htbp]
        \includegraphics[width=1\columnwidth]{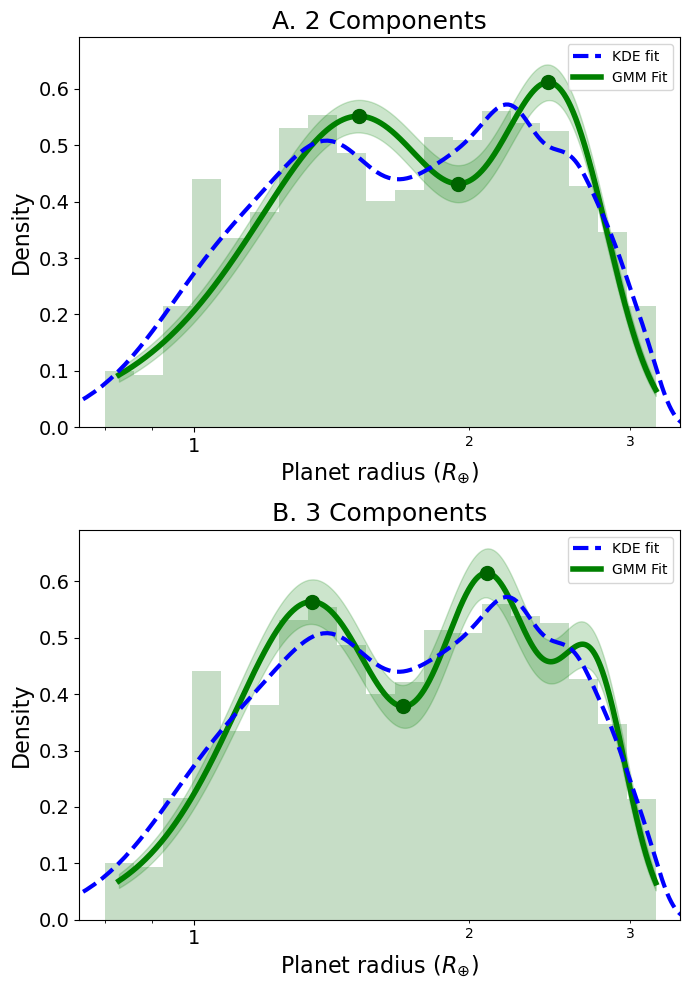}
        \caption{Comparison between the 2- and 3- component fit along with the KDE fits. It is clearly observed that the KDE fit goes along better with the 3- component GMM fit.}
        \label{kde+gmm}
    \end{figure}

For the GMM fitting, we tested various covariance structures (full, tied, diagonal, spherical) and found that the ‘tied’ covariance yielded the best fit. We then needed to determine the appropriate number of components. Given the known presence of both Super-Earth and Mini-Neptune populations, we initially used a 2-component model. However, this configuration failed to accurately capture the Mini-Neptune peak and tended to overestimate the location of the radius valley. Introducing a third component improved the fit significantly, suggesting that the Mini-Neptune population may not follow a simple Gaussian distribution \citep{Dattilo2024_cliff}. The slope values changed only marginally between the 2- and 3-component fits, supporting the robustness of the results. The derived slope values are presented in Table \ref{gmm}, and Figure \ref{2and3} illustrates that both 2- and 3-component models preserve the linear mass-radius scaling across GKM spectral types. Nonetheless, we adopt the 3-component model as it shows better agreement with the KDE (Figure \ref{kde+gmm}) and yields a lower Akaike Information Criterion (AIC), indicating superior predictive performance.\\
 
It is also worth noting that incorporating radius uncertainties into the GMM bootstrapping increased the error by only 0.0216. This modest increase suggests that statistical uncertainties dominate over radius measurement errors, further supporting the robustness of our analysis.\\



\section{Confirmed and Candidate Planets}
\label{planets_app}
\begin{figure}[htbp]
        \includegraphics[width=1\columnwidth]{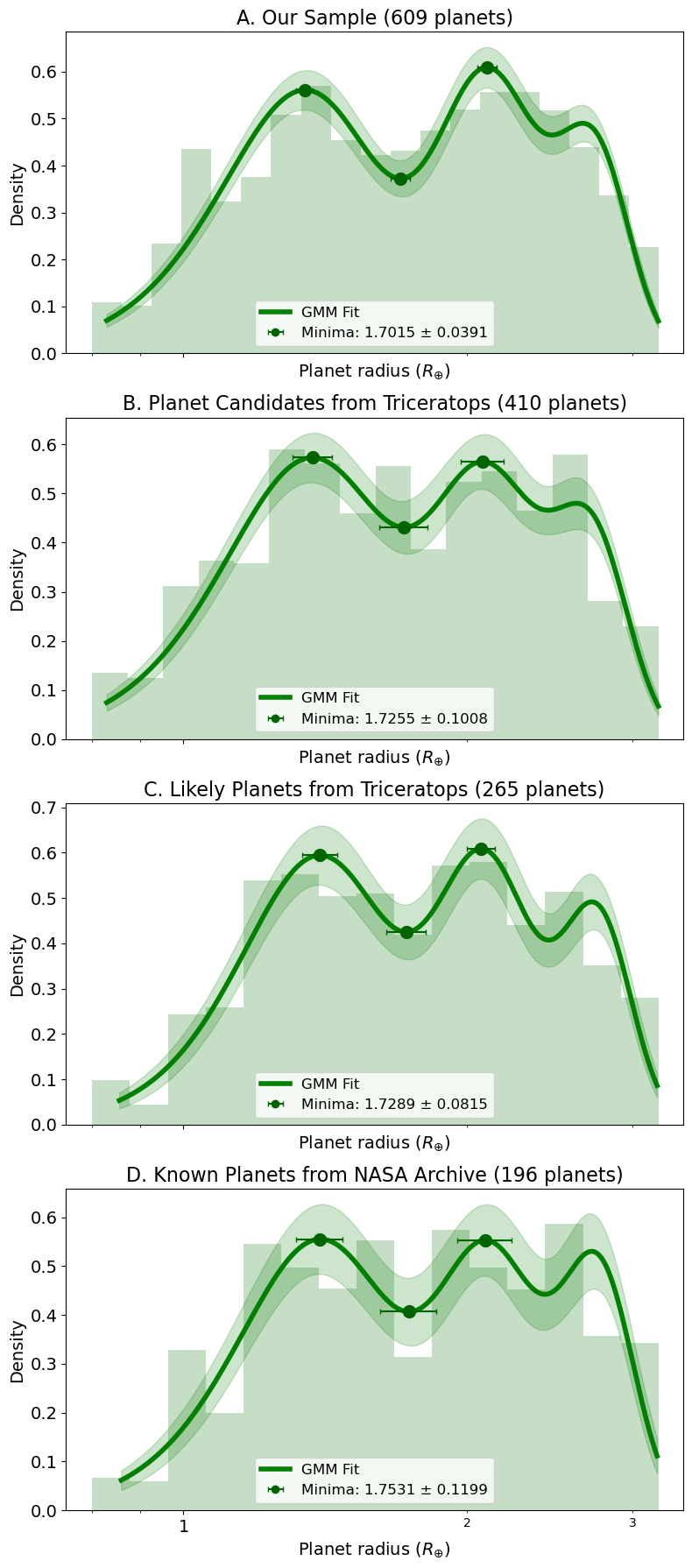}
        \caption{We performed our GMM analysis on various planet samples, from top, all planet candidates and confirmed planets from our sample (A), planet candidates (B) and likely planets (C) obtained from our Triceratops run \citep{code-tri} and confirmed planets from the NASA exoplanet archive (D) with number of planets in parentheses. As it is shown, including the planet candidates not only increases the planet sample; it also helps to reduce the error bars on the peaks and minima and precisely locate the radius valley.}
        \label{candvsconf}
    \end{figure}

\end{appendix}
\end{document}